\documentclass[11pt,twoside]{article}

\usepackage{asp2006}
\usepackage{epsf}
\usepackage{psfig}
\usepackage{lscape}
\usepackage{graphicx}

\begin{document}
\title{Investigation of the Faraday Rotation Measure and Magnetic Field Structures of Several AGN Jets on Sub-Parsec to Parsec Scales}   
\author{Shane P. O'Sullivan and Denise C. Gabuzda}   
\affil{Department of Physics, University College Cork, Cork, Ireland}    

\begin{abstract} 
Preliminary VLBA polarisation results on 6 ``blazars'' from 6.5-cm to 7-mm are presented here. Observing at several different wavelengths,
separated by short and long intervals, enabled reliable information about the magnetic (B) field structure to be
obtained and for the effect of Faraday Rotation to be determined and corrected. For all sources the magnitude
of the core Rotation Measure (RM) derived from the shorter wavelength data was greater than that derived from the longer wavelength data, consistent with a higher electron density and/or B-field strength closer to the central
engine. A transverse RM gradient was detected in the jet of 0954+658, providing evidence for the presence of a
helical B-field surrounding the jet. The RM in the core region of 2200+420 (BL Lac) displays sign changes in
different wavelength intervals (on different spatial scales); we suggest an explanation for this in terms of modest
bends in a helical B-field surrounding the jet.
\end{abstract}


\section{Introduction}
The Faraday effect causes a rotation of the plane of linear polarisation, described by: $\Delta\chi = RM \lambda^2$,
with the rotation measure (RM) determined by the integral of the electron density and the dot product of the
magnetic (B) field and the path length along the line of sight (LoS). A positive/negative RM indicates that the LoS
B-field is pointing towards/away from the observer. $\Delta\chi$ is the change in the electric vector position angle
(EVPA).
Previous results indicated the presence of different RM signs in the core regions of 6 blazars in different wavelength
intervals (O'Sullivan \& Gabuzda 2006). This has two main possible origins: (1) the LoS B-field changes with distance from the
centre of activity, or (2) since the previous observations were not simultaneous, it could be due to an
intrinsic change in the overall jet B-field structure between observing epochs. Our new 8 wavelength observations are designed to test these possibilities.

\section{Observations}
Multi-wavelength (6.5, 5.9, 3.8, 3.4, 2.3, 1.9, 1.3 cm \& 7 mm), simultaneous polarisation observations of 6
``blazars'' (0954+658, 1156+295, 1418+546, 1749+096, 2007+777, 2200+420) were obtained on the VLBA over a 24-hr period on 2 July 2006. The calibration and imaging were done in AIPS using standard techniques. The integrated (Galactic) RMs were subtracted, to better isolate the RM distribution in the immediate vicinity of the AGNs (Pushkarev 2001 and references
therein). Matched-resolution images were constructed for two seperate wavelength intervals using the 5.9-cm beamsize for the longer wavelength interval and the 1.9-cm beamsize for the shorter wavelength interval. This was done to obtain more reliable information about the RM distribution and B-field in the long and short wavelength regimes, as can be seen from the plot of $\chi$ vs. $\lambda^2$ for 1156+295 for the whole wavelength range in Figure 1.

\section{Rotation Measure Results}
A summary of preliminary results is displayed in Table 1. Seperate RMs were found for long and short wavelength intervals between which a clear transition was present (eg. Fig.1). The redshift, z, and the orientation of the jet EVPA relative to the jet direction ($\parallel$ aligned, $\perp$ orthogonal, $-$ no jet polarisation) are also given. For all sources the magnitude of the core RM is higher at shorter wavelengths, consistent with an increased electron number density closer to the central engine. A larger B-field strength can also contribute to this increase.

\begin{table}[!hb]
\caption{Summary of RM results. All values in $rad/m^2$}
\smallskip
\begin{center}
{\small
\begin{tabular}{ccccc}
\tableline
\noalign{\smallskip}
Blazar & z & Jet EVPA vs.& Core RM & Core RM \\
 & &Jet Direction& (Long $\lambda$ range)& (Short $\lambda$ range) \\
\noalign{\smallskip}
\tableline
\noalign{\smallskip}
$0954+658$ & $0.368$ & $\parallel$ & $-41\pm14$ & $-2207\pm386$\\
\noalign{\smallskip}
$1156+295$ & $0.729$ & $\parallel$ & $+136\pm6$ & $+1647\pm209$\\
\noalign{\smallskip}
$1418+546$ & $0.152$ & $\perp$ & $-65\pm22$ & $-483\pm55$\\
\noalign{\smallskip}
$1749+096$ & $0.320$ & $-$ & $-33\pm24$ & $-500\pm347$\\
\noalign{\smallskip}
$2007+777$ & $0.342$ & $\parallel$ & $+636\pm110$ & $+1946\pm140$\\
\noalign{\smallskip}
$2200+420$ & $0.069$ & $\parallel$ & $+746\pm76/-125\pm30$ & $-1144\pm36$\\
\noalign{\smallskip}
\tableline
\end{tabular}
}
\end{center}
\end{table}

 \begin{figure}[!ht]
  \begin{minipage}[t]{7.0cm}
  \begin{center}
  \includegraphics[width=6.5cm,clip]{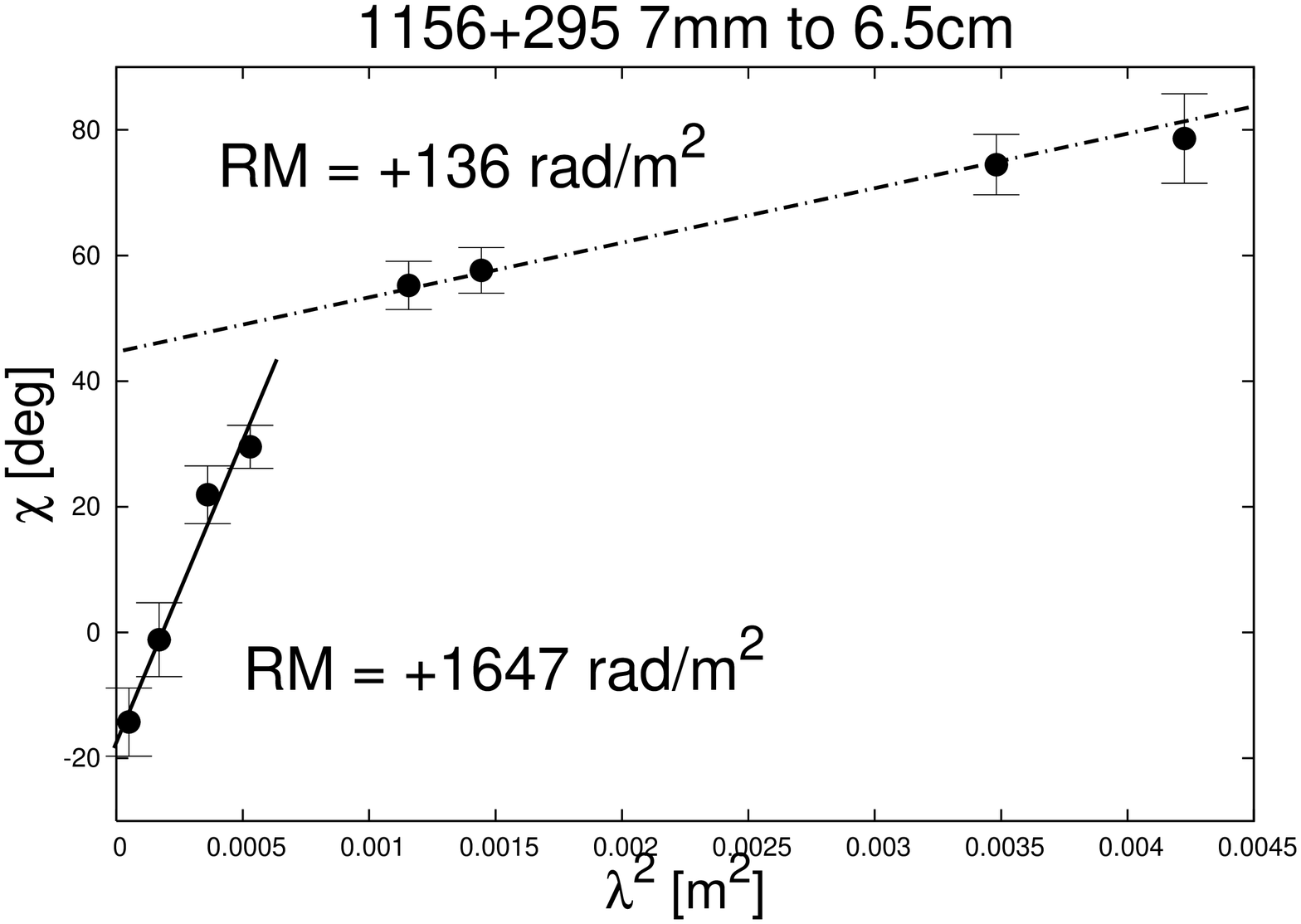}
  \end{center}
  \end{minipage}
  \hfill
   \begin{minipage}[t]{5.5cm}
   \begin{center}
   \includegraphics[width=4.0cm,clip]{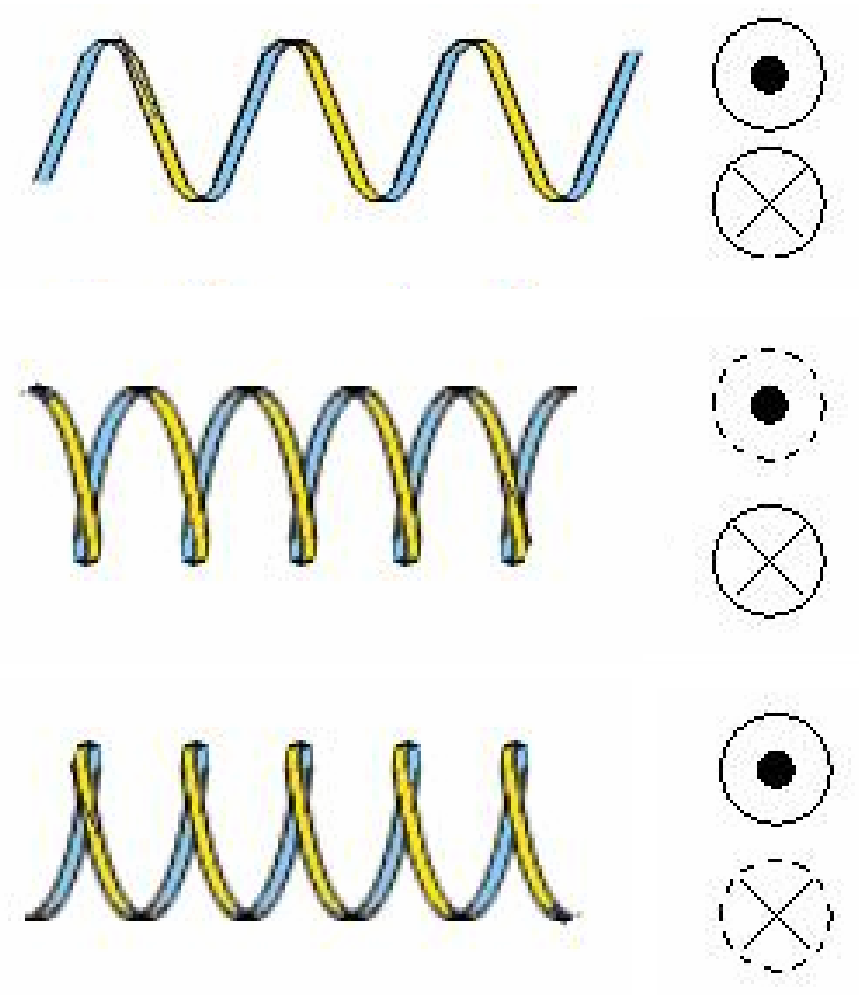}
   \end{center}
   \end{minipage}
  \caption{Left: Plot of $\chi$ vs. $\lambda^2$ for all observed wavelengths for 1156+295; the obvious increase in the slope at short wavelenghts shows the need to split up the wavelength range to obtain more reliable information about the RM and B-field on different scales. Right: Helical field viewed: Side-on (top), Tail-on (middle), Head-on (bottom). Solid and dashed circles indicate the relative strength of the RM on the two sides of the jet. A black dot/cross indicates that the LoS B-field is pointing towards/away from us.}
  \end{figure}

 \begin{figure}
  \begin{minipage}[t]{13.4cm}
  \begin{center}
  \includegraphics[width=10.5cm,clip]{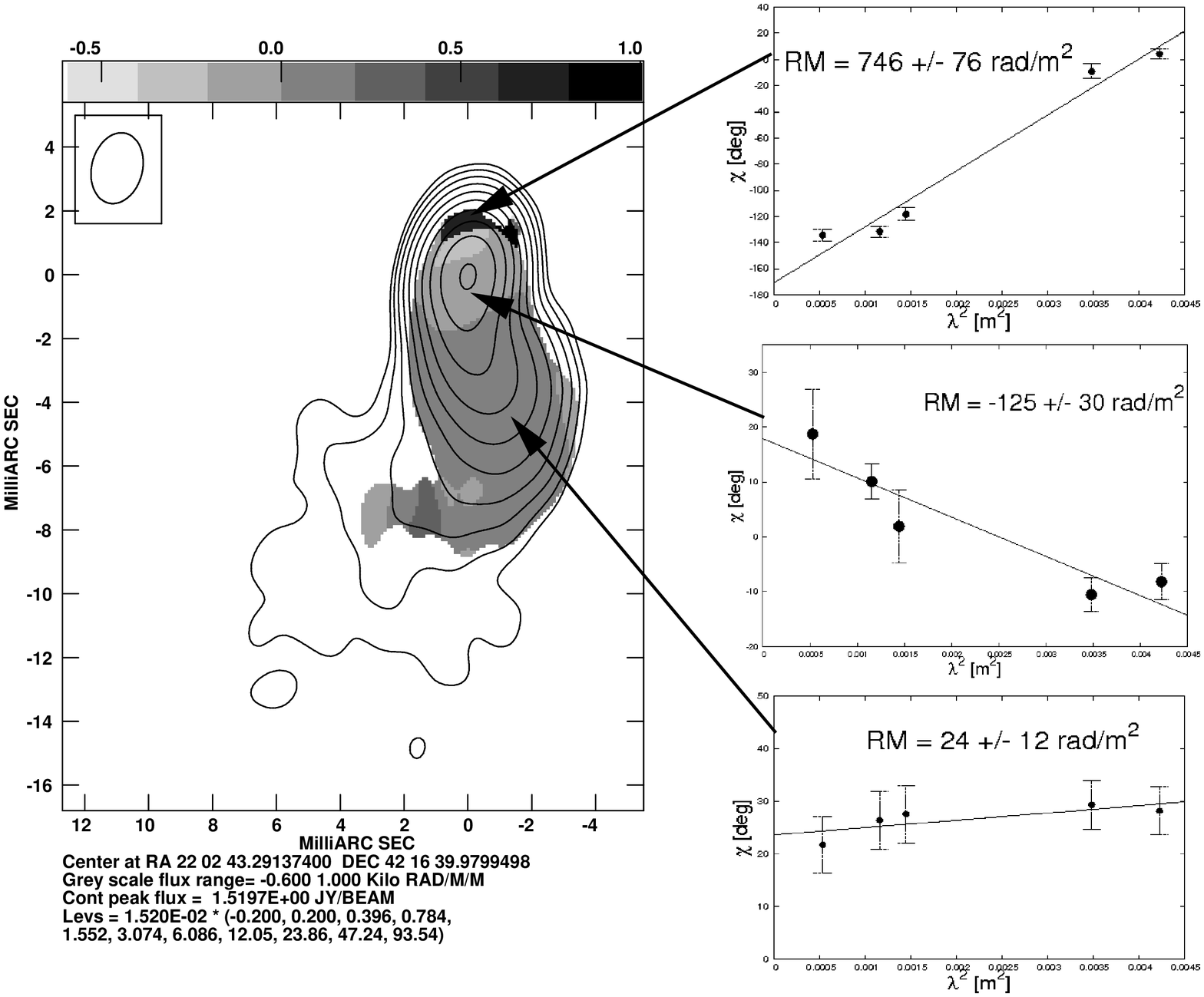}
  \caption{RM map for 2200+420 from 6.5-cm to 2.3-cm superimposed on the 5.9 cm \emph{I} map. The accompanying plots are of $\chi$ vs. $\lambda^2$ for the indicated regions of the source.}
  \end{center}
  \end{minipage}
  \end{figure}

 \begin{figure}
  \begin{minipage}[t]{13.4cm}
  \begin{center}
  \includegraphics[width=10.5cm,clip]{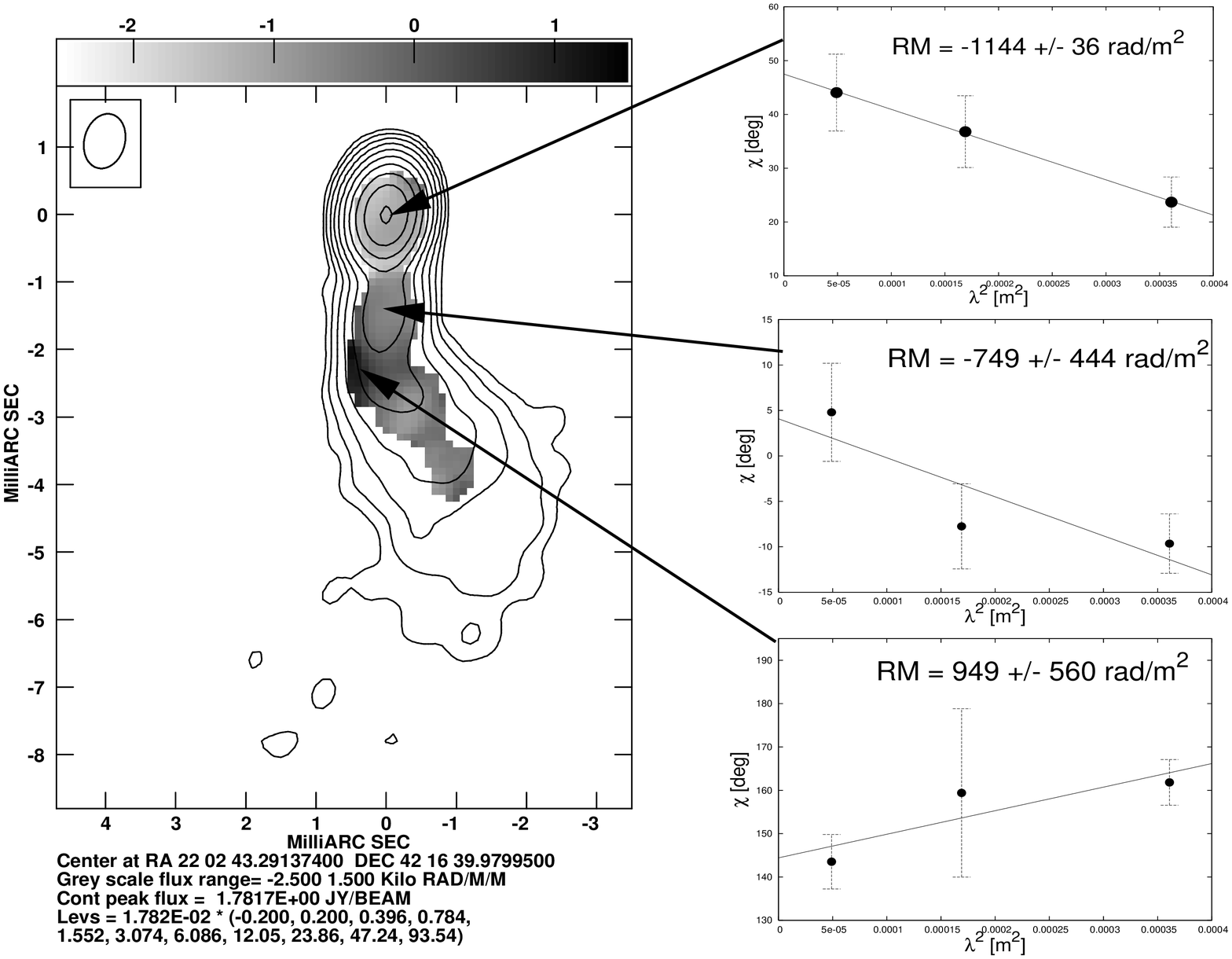}
  \caption{RM map for 2200+420 from 7-mm to 1.9-cm superimposed on the 1.9 cm \emph{I} map. The accompanying plots are of $\chi$ vs. $\lambda^2$ for the indicated regions of the source.}
  \end{center}
  \end{minipage}
  \end{figure}

\begin{figure}
 \begin{minipage}[t]{13.4cm}
 \begin{center}
 \includegraphics[width=10.0cm,clip]{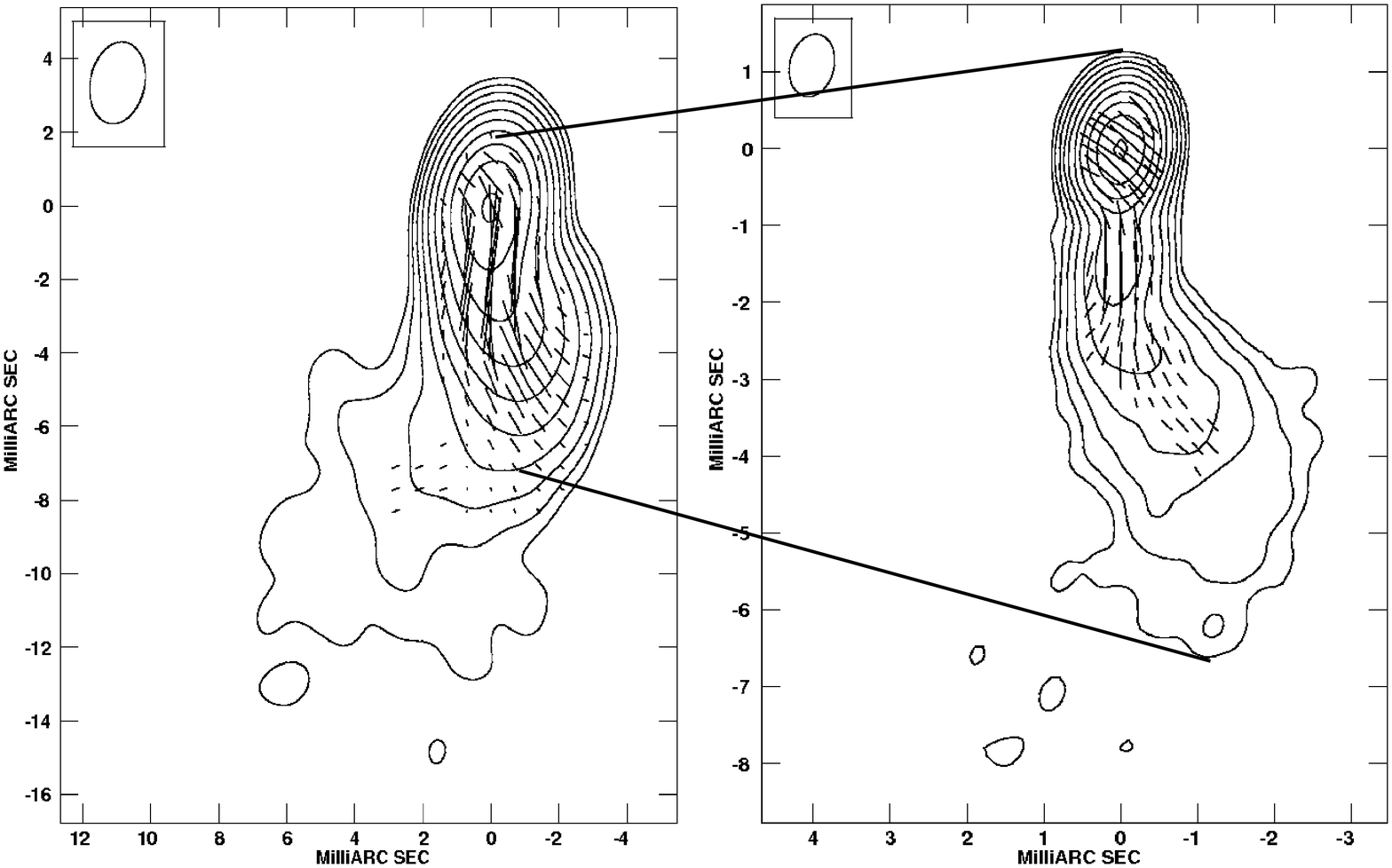}
  \caption{Left: 5.9 cm \emph{I} map, Right: 1.9cm \emph{I} map, for 2200+420; both with Faraday corrected EVPAs.
  Note alignment of jet EVPAs even as jet bends.
  }
 \end{center}
 \end{minipage}
 \end{figure}

2200+420 (BL Lac) has quite a complicated and variable structure. The inner-jet has changed from a southwesterly
direction in our 7 August 2002 observations (Gabuzda et al. 2006) to directly southwards in our current observations, consistent with the
``precessing nozzle'' proposed by Stirling et al. (2003). The RM map from 6.5-cm to 2.3-cm (Fig.2) shows
a large positive value of $+746$ $rad/m^2$ at the northern end, which we consider to be the true core, since our
spectral index map (corrected for the frequency dependent core shift; see Lobanov 1998) shows this region as being
most optically thick; a negative RM of $-125$ $rad/m^2$ is observed just south of this region, while the rest of the jet
has a low RM (essentially equal to zero within the errors) consistent with a much lower electron density in the vicinity of the optically thin jet.
The RM map of 2200+420 from 7-mm to 1.9-cm (Fig.3) displays an RM of $-1144$ $rad/m^2$ in the core, which is larger in
magnitude and different in sign than the observed northernmost RM for the longer wavelength data. The inner-jet RM has a
smaller value of $-732$ $rad/m^2$; these regions probably correspond to the region of negative RM near the core in the longer wavelength RM map (Fig.2). An interesting feature in Fig.3 is the high, positive RM detected at the eastern edge of the jet in the region where the it
bends, possibly indicating interation with the surrounding medium.

The fact that the Faraday corrected polarisation vectors for 2200+420 from both frequency intervals (Fig.4) remain well
aligned with the jet even as it goes through substantial bending can be understood if the implied transverse B-field
represents the toroidal component of a helical field. 1418+546 is the only source in this sample with jet polarisation perpendicular to the jet
direction (Table 1). This behaviour of the jet EVPAs is also natural if the jets have helical B-fields (Lyutikov et al. 2005), where
polarisation perpendicular to the jet direction occurs when the poloidal component of the helical B-field dominates.

A transverse RM gradient was detected in the jet of 0954+658, which is a strong signature of the presence of a helical
B-field surrounding the jet, confirmed by the results of Mahmud \& Gabuzda 2007 (these proceedings).

\section{Discussion}
Our results for 2200+420 confirm the presence of an RM sign reversal in the core region. Since the dominant jet B-field
is transverse to the jet and remains transverse while the jet bends, we will suppose a helical B-field surrounds the
jet. The observed RM sign reversal can be explained by a slight bend of the jet, due, for example, to a collision with material in the
parent galaxy or some instabilities inherent in the jet itself. (A longitudinal jet B-field with a change in the angle
to the line of sight could also cause a RM sign reversal, but this does not correspond to the observed B-field.)

A side-on view of a helical B-field (Fig.1 Top right) will have a RM that will be equally strong on both
sides of the jet, hence, a zero net RM will be observed for an unresolved jet. This would occur when the source is viewed at $1/\Gamma$ in the observer's frame.
For a tail-on view of a helical B-field (ie. $\theta > 1/\Gamma$) (Fig.1 Middle), the dominant RM will be from the bottom
half of the jet and a negative RM will be observed because the dominant LoS B-field will be
pointing away from us. (Assuming the jet is not fully resolved in the transverse direction.) Conversely, for a head-on
view of a helical B-field (ie. $\theta < 1/\Gamma$) (Fig.1 Bottom), a positive RM will be observed.

Therefore, regions with different RM signs in the jets of AGN can be explained within a helical B-field model as places
where the jet is observed at angles greater than or less than $~1/\Gamma$, due to bends in the jet. Since VLBI resolution
is usually not sufficient to completely resolve the true optically thick core, the VLBI ``core'' consists of emission
from the true core and some of the optically thin inner-jet. So if bends occur on scales smaller than the observed VLBI
core, ``core'' RMs with different signs could be derived from observations at different wavelengths (ie. probing
different scales of the inner-jet). In our future work, we will attempt to reconstruct the 3D path of the jet through
space using the combined information from the observed distributions of the total intensity, linear polarisation,
spectral index and rotation measure.


  \acknowledgements 
  Funding for this research was provided by the Irish Research Council for Science, Engineering and Technology.
  The VLBA is a facility of the NRAO, operated by Associated Universities Inc. under
  cooperative agreement with the NSF. 



\begin{thebibliography}{}
\bibitem[Gabuzda et al.(2006)]{Gab2006}
Gabuzda, D., Rastorgueva, E., Smith, P. \& O'Sullivan, S. 2006, MNRAS, 369, 1596

\bibitem[Lyutikov et al.(2005)]{Lyutikov2005}
Lyutikov, M., Pariev, V. I. \& Gabuzda, D. C. 2005, MNRAS, 360, 869

\bibitem[Lobanov(1998)]{Lob1998}
Lobanov, A. P. 1998, A\&A, 330, 79

\bibitem[Mahmud \& Gabuzda(2007)]{Mahmud2007}
Mahmud, M. \& Gabuzda, D. C. 2007, these Proceedings

\bibitem[O'Sullivan \& Gabuzda(2006)]{O'Sull2006}
O'Sullivan, S. P. \& Gabuzda, D. C., 2006, Proceedings of the 8th EVN Symposium

\bibitem[Pushkarev(2001)]{Pushk2001}
Pushkarev, A. B. 2001, Astron. Rep., 45, 667

\bibitem[Stirling et al.(2003)]{Stir2003}
Stirling, A. M. et al. 2003, MNRAS, 341, 405

\end{thebibliography}
\end{document}